\documentclass[proof]{pasj00}
\draft
\usepackage{pifont}
\usepackage{multirow}
\usepackage{lscape}
\usepackage{ccaption}

\begin{document}
\SetRunningHead{K. Torii et al.}{A Detailed Observational Study of Molecular Loops 1 and 2 in the Galactic Center}
\Received{2009/4/3}
\Accepted{xxxx/xx/xx}

\title{A Detailed Observational Study of Molecular Loops 1 and 2 in the Galactic Center}

\author{Kazufumi \textsc{Torii},\altaffilmark{1} 
        Natsuko \textsc{Kudo},\altaffilmark{1}
        Motosuji \textsc{Fujishita},\altaffilmark{1}
        Tokuichi \textsc{Kawase},\altaffilmark{1} \\
        Hiroaki \textsc{Yamamoto},\altaffilmark{1}
        Akiko \textsc{Kawamura},\altaffilmark{1}
        Norikazu \textsc{Mizuno},\altaffilmark{2}
        Toshikazu \textsc{Onishi},\altaffilmark{3} \\
        Akira \textsc{Mizuno},\altaffilmark{4} 
        Mami \textsc{Machida},\altaffilmark{1}
        Kunio \textsc{Takahashi},\altaffilmark{5}
        Satoshi \textsc{Nozawa},\altaffilmark{6} \\
        Ryoji \textsc{Matsumoto},\altaffilmark{7}
        and 
        Yasuo \textsc{Fukui}\altaffilmark{1}
        }

\altaffiltext{1}{Department of Astrophysics, Nagoya University, Chikusa-ku, Nagoya 464-8602}
\altaffiltext{2}{National Astronomical Observatory of Japan, Osawa, Mitaka, Tokyo 181-8588}
\altaffiltext{3}{Department of Physical Science, Osaka prefecture University, Sakai, Osaka 599-8531}
\altaffiltext{4}{Solar-Terrestrial Environment Laboratory, Nagoya University, Chikusa-ku, Nagoya 464-8601}
\altaffiltext{5}{Japan Agency for Marine-Earth Science and Technology, Kanazawa-ku, Yokohama, Kanagawa 236-0001, Japan}
\altaffiltext{6}{Department of Science, Ibaraki University, 2-1-1 Bunkyo, Mito, Ibaraki 310-8512}
\altaffiltext{7}{Faculty of Science, Chiba University, Inage-ku, Chiba 263-8522}

\email{torii@a.phys.nagoya-u.ac.jp, fukui@a.phys.nagoya-u.ac.jp}


\maketitle

\begin{abstract}
\citet{fuk2006} discovered two molecular loops in the Galactic center that are likely created by the magnetic flotation due to the Parker instability with an estimated field strength of $\sim$150 $\mu$G. Following the discovery, we present here a detailed study of the two loops based on NANTEN \atom{C}{}{12}\atom{O}{}{}($J$=1--0) and \atom{C}{}{13}\atom{O}{}{}($J$=1--0) datasets. 

The two loops are located in $l$ = 355$^\circ$ -- 359$^\circ$ and $b$ = 0$^\circ$ -- 2$^\circ$ at a velocity range of -20 -- -180 km s$^{-1}$. They have a projected total length of $\sim$ 600 pc and heights of $\sim$250 -- 300 pc from the Galactic disk at a distance of 8.5 kpc. They have loop-like filamentary distributions of 30 pc width and show bright foot points in the \atom{C}{}{12}\atom{O}{}{} emission in the edges of the loops at $b$ $\sim$ 0.8$^\circ$ -- 1.0$^\circ$. These foot points are characterized by velocity dispersions of 50 -- 100 km s$^{-1}$, much larger than those in the Galactic disk, supporting that the loops are located in the Galactic center within $\sim$1 kpc of Sgr A*. The loops also show large-scale velocity gradients of $\sim$30 -- 50 km s$^{-1}$ per $\sim$100 pc. 

We present an attempt to determine geometrical parameters and velocities of the loops by assuming that the loops having the same size are expanding and rotating at a constant radius $R$. The analysis yields that the loops are rotating at 50 km s$^{-1}$ and expanding at 150 km s$^{-1}$ at a radius of 670 pc from the center. 

We find some new features on smaller scales in the loops; first, the tops of loops 1 and 2 exhibit helical distributions in certain velocity ranges and, second, one of the foot point of loop 2 shows a sub-structure which mimics a smaller loop. These features possibly offer further indications for the strong magnetic field in the loops because of the similarity with the solar features resulting from magneto-hydrodynamic instabilities. 

The loops have counterparts of the \atom{H}{}{}\emissiontype{I} gas indicating that atomic gas also consists the magnetic loops. The \atom{H}{}{}\emissiontype{I} gas with weaker intensity fills the inner part of the loops where there is no significant \atom{C}{}{12}\atom{O}{}{} and \atom{C}{}{13}\atom{O}{}{} emission. The \atom{C}{}{12}\atom{O}{}{}, \atom{C}{}{13}\atom{O}{}{} and \atom{H}{}{}\emissiontype{I} datasets were used to estimate the total mass and kinetic energy of loops 1 and 2 to be $\sim5.5(\pm2.4) \times 10^6 \MO$ and $\sim$10$^{52}$ ergs, respectively. We also examined the far infrared distributions of IRAS toward the loops and found that the dust emission is well correlated with the gas distribution. The atomic hydrogen mass estimated from the dust emission is consistent with the above estimates.

Based on the new physical parameters, we present a reinforced argument that the huge size, velocity dispersions and that the kinetic energy are consistent with the magnetic floatation origin due to the Parker instability but is difficult to be explained by multiple stellar explosions. We also present detailed comparisons with the numerical simulations of magnetized nuclear disk by \citet{mac2009} and \citet{tak2009} and note that the model predictions show excellent agreements with the observations. 

\end{abstract}

\section{Introduction}
Molecular clouds are the sites of star formation and the distribution and dynamics of molecular gas should be crucial in the evolution of galaxies. The evolution of molecular gas may be significantly different in the central region of a galaxy from that in the disk because the stellar gravitational field is considerably stronger in the center due to increased stellar density. The strong stellar gravitational field makes the pressure much higher in the central region than in the disk and should influence considerably the gas dynamics. It is important to understand the physical conditions of molecular gas and their astronomical consequences in the Galactic center in our continuing efforts to elucidate galactic evolution.

Previous studies of the Galactic center show most of the molecular gas in the Galactic center is concentrated in the inner 300 pc which is called the gCentral Molecular Zoneh (hereafter CMZ, e.g., \cite{mor1996}). CMZ including Sgr A and Sgr B2 molecular clouds has been a region intensively studied in the Galactic center (e.g., \cite{fuk1977}; \cite{gus1983}; \cite{sco1974}). There are also molecular features with weaker intensities outside the CMZ up to nearly 1 kpc from the center. They include clump 1 ($l, b$ $\sim$ 355$^\circ$, 0$^\circ$), Clump 2 ($l, b$ $\sim$ 3$^\circ$, 0$^\circ$) and 5-deg feature ($l, b$ $\sim$ 5$^\circ$, 0$^\circ$) \citep{ban1977} and at least several weak \atom{C}{}{}\atom{O}{}{} emission features detected at a lower resolution of 8$'$.8 (\cite{dam1987}; \cite{bit1997}) although these features outside the CMZ received little attention so far. 

Fukui et al. (2006; hereafter Paper I) discovered that two molecular loops are located toward 355.5$^\circ < l < 358^\circ$ and 0$^\circ < b < 2^\circ$ by analyzing the NANTEN Galactic plane survey dataset of the \atom{C}{}{12}\atom{O}{}{}($J$=1--0) emission. These loops have heights of $\sim$ 2 degrees from the galactic plane and the projected lengths of 3 -- 4 degrees. The loops have two foot points that are brightest spots in the \atom{C}{}{}\atom{O}{}{} emission on each end at Galactic latitudes around 0.8 degrees. The velocity dispersions of the molecular gas in the loops are $\sim$50 -- 100 km s$^{-1}$ at the foot point of loops, which are characteristic to the molecular gas in the Galactic center and are much larger than that of the disk molecular clouds whose velocity widths are less than $\sim$ 10 km s$^{-1}$.@Most recently, \citet{fuj2009} have discovered loop 3 which is located in the same direction with loops 1 and 2 at a positive velocity range of $V_{\mathrm{LSR}}$ = 20 -- 200 km s$^{-1}$. Loop 3 shows a loop-like outer shape and similar height and length with loops 1 and 2 and it has two foot points with broad velocity dispersions. 

These three loops are interpreted by magnetic flotation caused by the Parker instability and Paper I presented the first observational evidence for the Parker's model in the galactic scale. The differential rotation in the nuclear disk creates toroidal magnetic field where the molecular gas is frozen-in to the magnetic field; the molecular gas in the disk is ionized at an ionization degree of 10$^{-8}$ -- 10$^{-7}$ by cosmic ray protons and the rate is high enough for the frozen-in condition that are described by the ideal magneto-hydrodynamics. The molecular gas layer in the disk is dynamically supported by the toroidal field against the stellar gravity in the $z$ direction. This configuration is unstable to disturbances and the gas rises up as a loop by the Parker instability \citep{par1966}. The gas inside the loop then flows down to the disk by the stellar gravity and the down flowing gas forms shock fronts above the disk \citep{mat1988}. This shock interaction forms large velocity dispersions at the foot points of loops, heating and compressing the gas by the shock fronts (Paper I). In particular, the most recent global numerical simulations that deal with the 2kpc-radius nuclear disk by \citet{mac2009} has revealed general properties of the magnetic floatation loops over the entire disk including that the one-armed non-axisymmetric (m = 1) mode tends to dominate the gas distribution. Subsequently, \citet{tak2009} carried out two-dimensional simulations of a local part of the Galactic center disk. They assumed a stratified disk consisting of cool layer ($T \sim 10^3$ K), warm layer ($T \sim 10^4$ K), and hot layer ($T \sim 10^5$ K) to confirm the observed galactic loops, and found that the flat-top loop efficiently accumulate the gas to make a dense layer at the top of the loop. 

In the CMZ, the molecular gas shows high temperatures and violent motions. The temperature is derived to be $\sim$30 -- 300 K using \atom{C}{}{}\atom{O}{}{}, \atom{N}{}{}\atom{H}{}{}$_3$, \atom{H}{}{}$_2$, \atom{H}{}{}$_3^+$ lines (\cite{mar2004}; \cite{hut1993}; \cite{rod2001}; \cite{oka2005}; \cite{nag2007}). The velocity dispersion is 15 -- 50 km s$^{-1}$, which is significantly larger than that in the molecular clouds in the Galactic disk (e.g., \cite{mor1996}; \cite{gus2004}). The causes of the high temperature and large velocity dispersion have been the longstanding puzzles in the last few decades. The magnetic field is as strong as 0.1 -- 1 mG as indicated by Radio Arc and other non-thermal filaments which are distributed within 1 degree of the center, often vertically to the Galactic plane (Yusef-Zadeh et al. 1984; 2004), and perhaps by the infrared double helix \citep{mor2006}. The magnetic flotation model has a potential to offer a coherent explanation of the origin of the high temperature and large velocity dispersion of molecular gas in the Galactic center as argued in Paper I because the high velocity motions are natural consequences of the gas motion along the magnetic loops which form shock fronts around them most significantly at the foot points.

Following the discovery of loops 1 and 2, it is important to clarify details of the loops including additional indications of the shock interaction. We here present a detailed analysis of loops1 and 2 by using the NANTEN Galactic plane survey, the GPS \atom{C}{}{12}\atom{O}{}{} and \atom{C}{}{13}\atom{O}{}{} datasets. Section 2 describes the \atom{C}{}{12}\atom{O}{}{} and \atom{C}{}{13}\atom{O}{}{}($J$=1--0) datasets. Detailed observational properties of the magnetic flotation loops are presented in section 3. Section 4 offers discussion and section 5 summarizes the paper. 

\section{NANTEN Datesets}
Basic parameters of the observations are summarized in Table \ref{tab:obs}.

\subsection{\atom{C}{}{12}\atom{O}{}{}}

\atom{C}{}{12}\atom{O}{}{}($J$=1--0) dataset toward the Galactic center region was taken with the NANTEN 4-m radio telescope at Las Campanas Observatory in Chile during the period from March 1999 to September 2001. The half-power beamwidth (HPBW) was 2$'$.6 at 115 GHz, the frequency of \atom{C}{}{12}\atom{O}{}{}($J$=1--0). The front end was a 4 K cryogenically cooled Nb superconductor-insulator-superconductor (SIS) mixer receiver \citep{oga1990} that provided a typical system temperature of $\sim$280 K in the single-side band, including the atmosphere toward the zenith. The spectrometer was an acousto-optical spectrometer (AOS) with 2048 channels. The frequency coverage and resolution were 250 MHz and 250 kHz, corresponding to a velocity coverage of 650 km s$^{-1}$ and a velocity resolution of 0.65 km s$^{-1}$, respectively, at 115 GHz. The intensity calibration was made by using the ambient temperature load. The absolute antenna temperature was calibrated by observing rho-Oph East [R.A.(1950) = \timeform{16h29m20.9s}, Dec.(1950) = \timeform{-24D22'13''}] every 2 hours, whose absolute temperature was assumed to be 15 K. The telescope pointing was measured to be accurate to within 20$''$ by radio observations of planets in addition to optical observations of stars with a CCD camera attached to the telescope. The observed region was 240 square degrees toward -12$^{\circ} < l < 12^{\circ}$ and -5$^{\circ} < b < 5^{\circ}$ at a grid spacing of 4$'$, corresponding to 10 pc at a distance of the Galactic center, 8.5 kpc. In total, 54,000 positions were observed. The integration time per point was 4 -- 9 s, resulting in typical r.m.s. noise fluctuations of 0.36 K at a velocity resolution of 0.65 km s$^{-1}$.

\subsection{\atom{C}{}{13}\atom{O}{}{}}

\atom{C}{}{13}\atom{O}{}{}($J$=1--0) dataset toward the Galactic center region was taken with the same instrument as the \atom{C}{}{12}\atom{O}{}{}($J$=1--0) observation in October 2003. The half-power beam width (HPBW) was 2$'$.7 at 110 GHz, the frequency of \atom{C}{}{13}\atom{O}{}{}($J$=1--0). The same receiver provided a typical system temperature of $\sim$100 K in the single-side band, including the atmosphere toward the zenith. The same spectrometer provided the frequency coverage and resolution, 250 MHz and 250kHz, corresponding to a velocity coverage of 620 km s$^{-1}$ and a velocity resolution of 0.62 km s$^{-1}$, respectively, at 110 GHz. The absolute antenna temperature was calibrated by observing rho-Oph East [R.A.(1950) = \timeform{16h29m20.9s}, Dec.(1950) = \timeform{-24D22'13''}] every 2 hours. Absolute radiation temperature of rho-Oph East was assumed to be 10 K. The observed region was 28 square degree toward -6$^{\circ}$ $<$ $l$ $<$ 8$^{\circ}$ and -1$^{\circ}$ $<$ $b$ $<$ 1$^{\circ}$ at a grid spacing of 2$'$. In total, 25,200 positions were observed. The integration time per point was 10 -- 15 s, resulting in typical r.m.s. noise fluctuations of 0.20 K at a velocity resolution of 0.62 km s$^{-1}$.

\section{Results}
\subsection{\atom{C}{}{}\atom{O}{}{} Distribution of Loops 1 and 2}

Figure \ref{fig:12CO.gcall.lb+lv} shows the large-scale distribution of \atom{C}{}{12}\atom{O}{}{} intensity in an area of $l$ = 350$^\circ$ -- 10$^\circ$ and $b$ = -5$^\circ$ -- 5$^\circ$ in $V_{\mathrm{LSR}}$= -300 -- 300 km s$^{-1}$.  Figure \ref{fig:12CO.gcall.lb+lv}a shows the total integrated intensity distribution of \atom{C}{}{12}\atom{O}{}{}. In Figure \ref{fig:12CO.gcall.lb+lv}a most of the emission along the galactic plane is the 3-kpc expanding arm and the CMZ is dominant in $l$ = 359$^\circ$ -- 2$^\circ$. Locations of loops 1 and 2 and loop 3 toward $l$ = 355$^\circ$ -- 359$^\circ$ are shown by boxes with broken lines in Figure \ref{fig:12CO.gcall.lb+lv}a and Figure \ref{fig:12CO.gcall.lb+lv}b. In Figure \ref{fig:12CO.gcall.lb+lv}b, the longitude velocity diagram, the molecular gas in the Galactic center region is seen as broad features whose velocity extents are a few tens km s$^{-1}$ or more. The narrow features near 0 km $s^{-1}$ are clouds outside the Galactic center and the 3-kpc arm is seen as a narrow feature at $\sim$-50 km s$^{-1}$ at $l$ = 0$^\circ$ with a velocity gradient of $\sim$8 km s$^{-1}$/deg. 

\begin{figure}
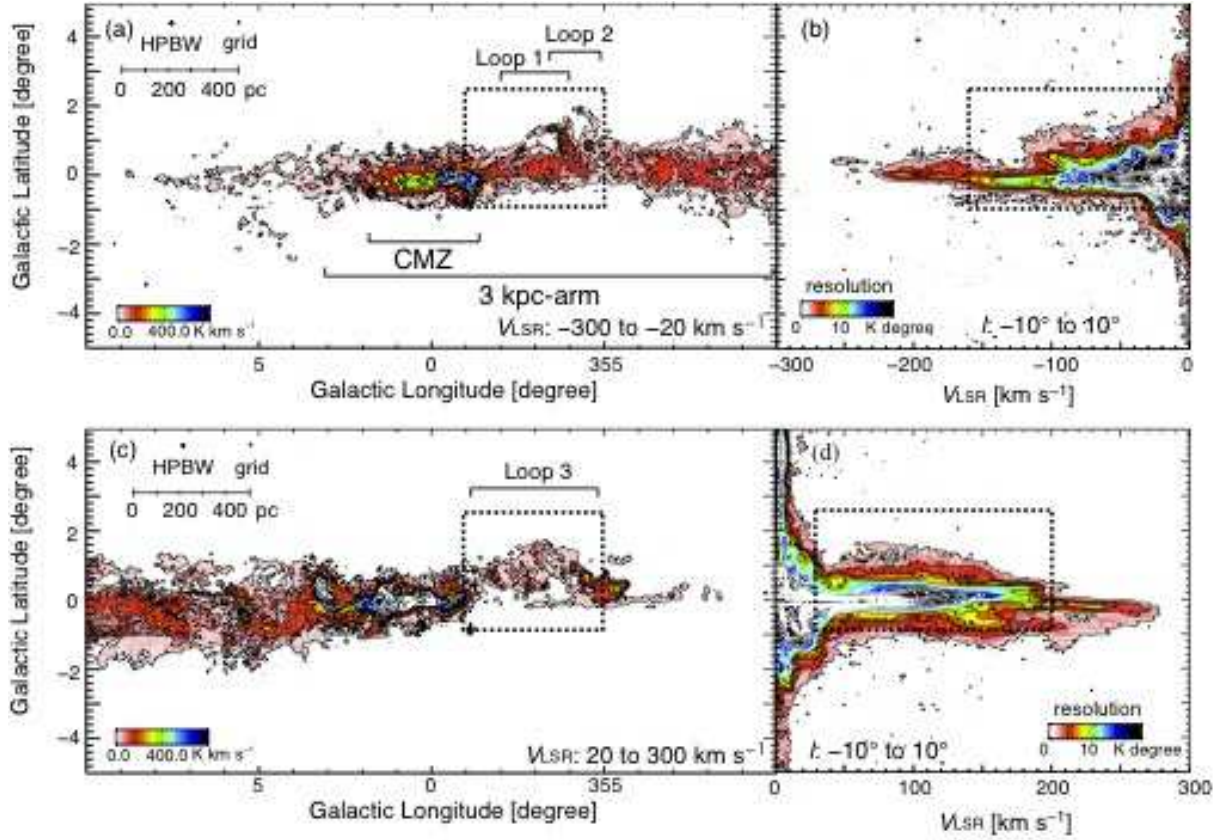

  \begin{center}
    \FigureFile(160.7mm, 261.05mm){12CO.gcall.loop123.lb+vb.eps2}
  \end{center}
  \caption{(a, b) Integrated intensity distributions of loops 1 and 2 in the Galactic center. Dotted boxes show the region that loops are located. (a) Velocity integrated intensity distributions. Integration range in velocity is from -300 to -20 km s$^{-1}$. Contours are plotted at 10, 20, 60, 100, 220, 340 K km s$^{-1}$ in black and 460, 820, 1180, 1540, 1900 K km s$^{-1}$ in white. (b) Velocity -- latitude diagram. Velocity resolutions are smoothed to 3 km s$^{-1}$. Integration range in galactic longitude is from -10$^\circ$ to 10$^\circ$. Contours are plotted every 1.6 K degree from 0.8 K degree (black) and then every 3.2 K degree from 6th one (white). (c, d) Integrated intensity distributions of loop 3 in the Galactic center. Dotted boxes show the region that loop is located. (c) Velocity integrated intensity distributions. Integration range in velocity is from 20 to 300 km s$^{-1}$. Contours are plotted at the same levels of figure (a). (d) Velocity -- latitude diagram. Velocity resolutions are smoothed to 3 km s$^{-1}$. Integration range in galactic longitude is from -10$^\circ$ to 10$^\circ$. Contours are plotted at the same levels of figure (b).}\label{fig:12CO.gcall.loop123.lb+vb}
\end{figure}

 Figure \ref{fig:12CO.gcall.loop123.lb+vb} shows the integrated intensity distributions for the negative and positive velocity ranges. Figure \ref{fig:12CO.gcall.loop123.lb+vb}a and \ref{fig:12CO.gcall.loop123.lb+vb}b show the \atom{C}{}{12}\atom{O}{}{} distributions in the negative velocity range in the $l$-$b$ plane and $b$-$v$ plane, respectively, and loops 1 and 2 are indicated in Figure \ref{fig:12CO.gcall.loop123.lb+vb}a.  Figure \ref{fig:12CO.gcall.loop123.lb+vb}c and \ref{fig:12CO.gcall.loop123.lb+vb}d show the \atom{C}{}{12}\atom{O}{}{} distributions in the positive velocity range in the $l$-$b$ plane and $b$-$v$ plane, respectively, and loop 3 is indicated in Figure \ref{fig:12CO.gcall.loop123.lb+vb}c.  
 
 Figure \ref{fig:Loop1.lblv+13CO.2} shows close-up images of \atom{C}{}{12}\atom{O}{}{} and \atom{C}{}{13}\atom{O}{}{} for loops 1 and 2 for two velocity ranges, from -180 to -90 km s$^{-1}$ and from -90 to -40 km s$^{-1}$, in an area of $l$ = 355$^\circ$ -- 359$^\circ$ and $b$ = 0$^\circ$ -- 2$^\circ$ shown by a rectangle in Figure \ref{fig:12CO.gcall.loop123.lb+vb}a. Figures \ref{fig:Loop1.lblv+13CO.2}a-d are for loop 1; Figures \ref{fig:Loop1.lblv+13CO.2}a and \ref{fig:Loop1.lblv+13CO.2}b are the $l$-$b$ diagrams of \atom{C}{}{12}\atom{O}{}{} and \atom{C}{}{13}\atom{O}{}{}, respectively, and Figures \ref{fig:Loop1.lblv+13CO.2}c and \ref{fig:Loop1.lblv+13CO.2}d are the $l$-$v$ diagrams of \atom{C}{}{12}\atom{O}{}{} and \atom{C}{}{13}\atom{O}{}{}, respectively. Figures \ref{fig:Loop1.lblv+13CO.2}e-h are for loop 2; Figures \ref{fig:Loop1.lblv+13CO.2}e and \ref{fig:Loop1.lblv+13CO.2}f are the $l$-$b$ diagrams of \atom{C}{}{12}\atom{O}{}{} and \atom{C}{}{13}\atom{O}{}{}, respectively, and Figures \ref{fig:Loop1.lblv+13CO.2}g and \ref{fig:Loop1.lblv+13CO.2}h are the $l$-$v$ diagrams of \atom{C}{}{12}\atom{O}{}{} and \atom{C}{}{13}\atom{O}{}{}, respectively.   
	
Figure \ref{fig:Loop1.lblv+13CO.2}a shows that loop 1 is a remarkable filamentary feature of $\sim$30 pc width and $\sim$300 pc projected length, being elevated from the plane by $\sim$200 pc. These parameters are all assumed the distance to the Galactic center as 8.5 kpc. It shows a bright spot at ($l, b$) $\sim$ (356.0$^\circ$, 1.0$^\circ$), which has strong intensity gradients toward the plane and also toward the east. We also recognize a less enhanced peak toward ($l, b$) $\sim$ (357.4$^\circ$, 0.8$^\circ$). We shall call these two peaks gfoot pointsh following Paper I. Figure \ref{fig:Loop1.lblv+13CO.2}c shows that these two foot points have enhanced extents of $\sim$50 km s$^{-1}$ or more. The \atom{C}{}{13}\atom{O}{}{} emission is significant only toward the foot point at (356.0$^\circ$, 1.0$^\circ$) while the observations does not cover the region above $b$ = 1$^\circ$.
	
Figure \ref{fig:Loop2.lblv+13CO.2}e shows that loop 2 is an outstanding feature with a width of $\sim$30 pc and a length of $\sim$400 pc and is elevated by $\sim$300 pc from the plane. A feature at b $\sim$ 1.4$^\circ$ extending toward the center is part of loop 1. Loop2 shows two bright spots with enhanced extents of $\sim$60 km s$^{-1}$ toward ($l, b$) $\sim$ (355.4$^\circ$, 0.8$^\circ$) and (356.1$^\circ$, 0.8$^\circ$). The \atom{C}{}{13}\atom{O}{}{} emission is detected toward these two spots (Figure \ref{fig:Loop2.lblv+13CO.2}c,d) although the coverage is limited to $b$ less than 1$^\circ$. We shall call these two spots gfoot pointsh of loop 2 (Paper I). 
	
Figure \ref{fig:specloop12.new} shows typical \atom{C}{}{12}\atom{O}{}{} and \atom{C}{}{13}\atom{O}{}{} line profiles toward loops 1 and 2. These show the line profiles are not Gaussian but have shapes of peaked triangles. We note that the profiles toward the foot points show asymmetry with a strong skew. The velocity span is as large as 40 km s$^{-1}$ towards the loop top and becomes even larger as 100 km s$^{-1}$ towards the foot points.  

Figures \ref{fig:loopallvel2}a and \ref{fig:loopallvel2}b show distributions of the averaged velocity in the loops where the averaging was made in a 5 arcmin Gaussian beam. The larger beam size was taken in order to suppress small scale irregularities. We roughly estimate velocity gradients of $\sim$0.2 km s$^{-1}$ pc$^{-1}$ and $\sim$0.35 km s$^{-1}$ pc$^{-1}$ along the loops for loops 1 and 2, respectively.

In order to show more details of the velocity distributions we show a series of velocity channel distributions every 10 km s$^{-1}$ as 12 panels in Figure \ref{fig:12co.channel.01}. We identify the foot points of loops 1 and 2 towards $l$ = 355$^\circ$, 357$^\circ$ and 359$^\circ$ $b$ = 0.8$^\circ$. From -120 to -90 km s$^{-1}$ we find a vertical feature toward $l$ = 356$^\circ$ at $b$ from 1.4$^\circ$ to 1.6$^\circ$. This feature between loops 1 and 2 does not seem to be part of the loops. 

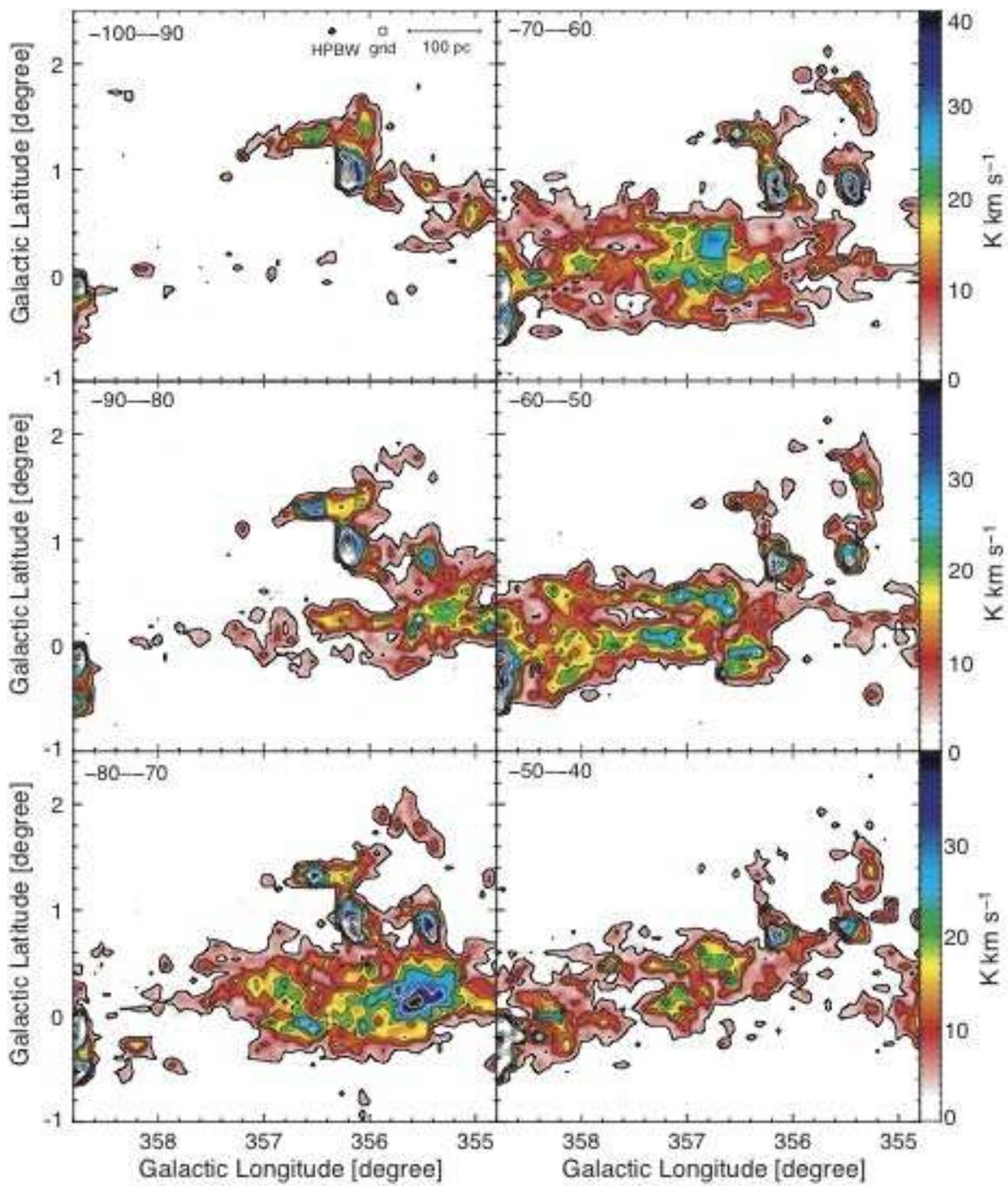
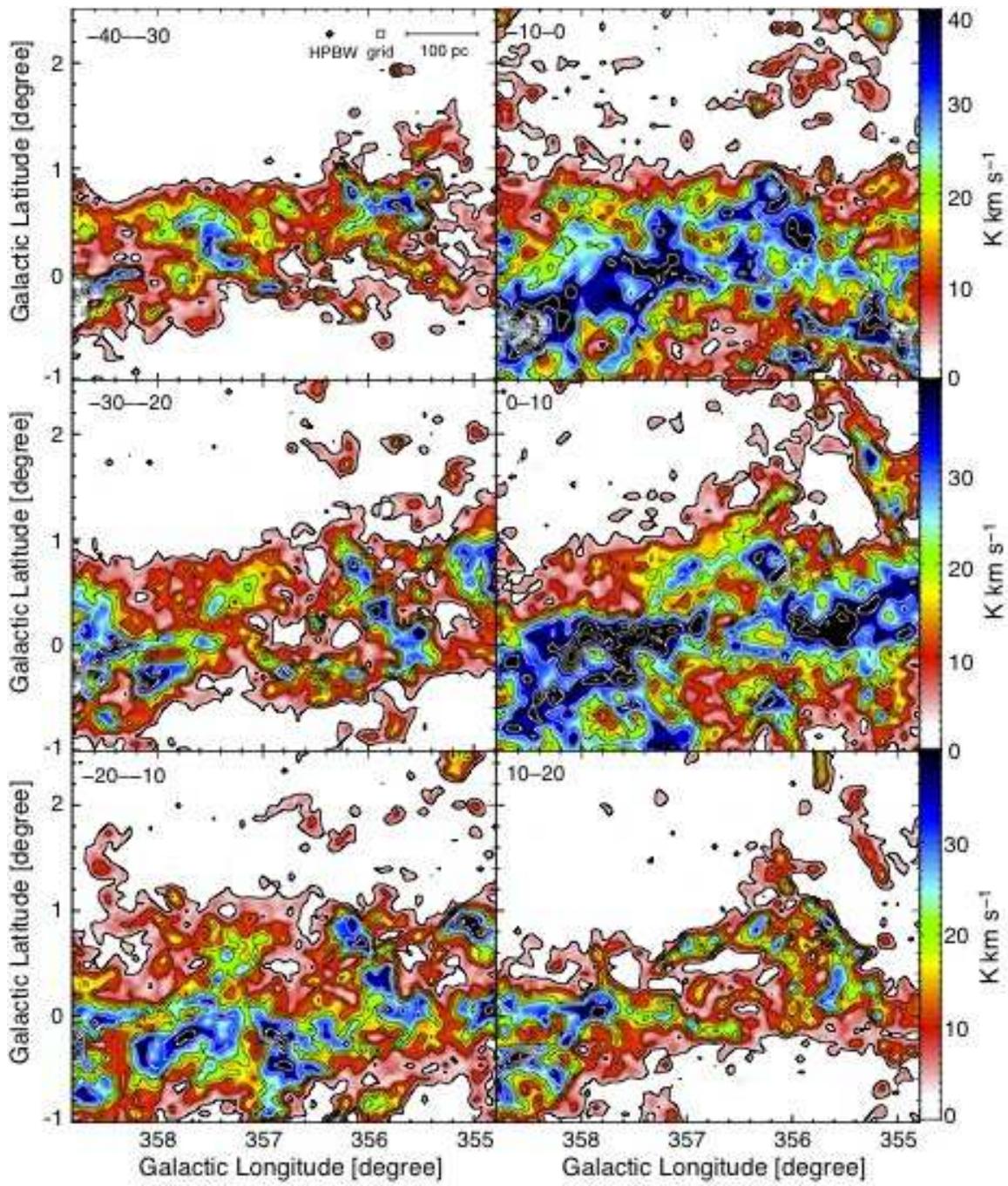
\begin{figure}
  \begin{center}
    \FigureFile(139.37mm,187.82mm){helix_a.eps2}
  \end{center}
  \caption{Velocity-integrated intensity map helical feature in loop 1. (a) Velocity integrated intensity map in \atom{C}{}{12}\atom{O}{}{}($J$=1--0). The integration range is from -129 to -107 km/s. Contours are plotted every 2 K km/s from 2.6 K km/s. Black box indicates the region shown in figure (b). (b) Velocity channel maps of helical feature integrated over 4 km/s with interval of 2 km/s. Contours are plotted every 0.8 K km/s from 1.6 K km/s. Dotted lines show the fitting result with the parameters shown in Table \ref{tab:helix}.}\label{fig:helix_a}
\end{figure}

In addition, we note three regions show peculiar distributions that are not part of simple loops as shown in Figures \ref{fig:helix_a}, \ref{fig:helix_b} and \ref{fig:helix_c}. 

Figure \ref{fig:helix_a} shows the details of the top of loop 1 every 4 km s$^{-1}$ from -129 to -107 km s$^{-1}$. The panels from -121 to -113 km s$^{-1}$ show S-shaped distributions in a range $l$ = 357.0$^\circ$ -- 357.4$^\circ$ and those from -115 to -113 km s$^{-1}$ also show a similar feature in a range from (356.8$^\circ$ - 357.3$^\circ$, 0.9$^\circ$).

Figure \ref{fig:helix_b}  shows the top of loop 2 every 4 km s$^{-1}$ from -86 to -66 km s$^{-1}$. A peculiar peaked shape is found from $l$ = 355.4$^\circ$ to 355.8$^\circ$. and $b$ = 1.8$^\circ$ to 2.1$^\circ$. Figure \ref{fig:helix_c} shows the area in the north of the foot point. The panels from -50 to -42 km s$^{-1}$ show a vertical feature at $l$ = 355.6$^\circ$ and the panels from -44 to -36 km s$^{-1}$ indicate a bridge-like feature that connects this vertical feature and the local intensity peak at ($l, b$)=(355.0$^\circ$ -- 355.2$^\circ$) in panel from -44 to -40 km s$^{-1}$. In addition, the panel from -34 to -30 km s$^{-1}$ shows another vertical feature at $l$ = 355.2$^\circ$ and $b$ = 1.4$^\circ$ to 1.8$^\circ$.

\subsection{Comparison with \atom{H}{}{}\emissiontype{I}}

Figure \ref{fig:LBLV.CO+HI} shows a comparison between \atom{C}{}{12}\atom{O}{}{} and \atom{H}{}{}\emissiontype{I} where the \atom{H}{}{}\emissiontype{I} data were taken with the Parkes 64m telescope having 16 arcmin beam \citep{mcc2005}. Figures \ref{fig:LBLV.CO+HI}a and \ref{fig:LBLV.CO+HI}b are for loop 1 and Figures \ref{fig:LBLV.CO+HI}c and \ref{fig:LBLV.CO+HI}d for loop 2 in the $l$-$b$ and $l$-$v$ diagrams. The \atom{H}{}{}\emissiontype{I} distribution seems similar to \atom{C}{}{12}\atom{O}{}{} in the both loops indicating that the loops accompany atomic components also, while the \atom{H}{}{}\emissiontype{I} seems enveloping the CO. It is notable that the inside of loop 1 shows weak \atom{H}{}{}\emissiontype{I} emission (Figure \ref{fig:LBLV.CO+HI}a). Loop 2 is not so clearly recognized as loop 1 but the western half of loop 2 shows a possible \atom{H}{}{}\emissiontype{I} counterpart in Figure \ref{fig:LBLV.CO+HI}b. The outside of loops 1 and 2 shows little sign of \atom{H}{}{}\emissiontype{I} emission. Figure \ref{fig:BV.CO+HI} shows $b$-$v$ cuts of \atom{C}{}{12}\atom{O}{}{} and \atom{H}{}{}\emissiontype{I}. Loop 1 is clearly seen from v $\sim$ -150 to -100 km s$^{-1}$ around $b$ $\sim$ 1.0 (Figure \ref{fig:BV.CO+HI}a) with the \atom{H}{}{}\emissiontype{I} associated with the CO (Figure \ref{fig:BV.CO+HI}b).  

\subsection{Comparison with Dust emission}
In order to test if the loops in CO and \atom{H}{}{}\emissiontype{I} are seen in the dust emission we have compared dust emission obtained with IRAS at 60 and 100 $\mu$m, where we did not use shorter wavelength data because of the possible contamination by the zodiacal light. 
 Figures \ref{fig:IRAS60100+CO.total.lb}a and \ref{fig:IRAS60100+CO.total.lb}b show comparison between the total integrated intensity of \atom{C}{}{12}\atom{O}{}{} from -300 to 300 km s$^{-1}$, including loops 1, 2 and 3, and IRAS 60 $\mu$m and 100 $\mu$m emissions, respectively. We find that the enhanced CO emission above $b$ = 1$^\circ$ in a range of $l$ = 355$^\circ$ -- 358$^\circ$ shows a good coincidence with the dust emissions; the hole of the CO emission at $l$ = 356.5$^\circ$ and the vertical feature at $l$ = 355.3$^\circ$, the western part of loop 2, are recognized well particularly at 100 $\mu$m.  Figures \ref{fig:IRAS60100+HI.total.lb}a and \ref{fig:IRAS60100+HI.total.lb}b, a similar comparison with the \atom{H}{}{}\emissiontype{I} convolved to the same beam size, show that the dust emissions show similar elevated distribution as in Figure \ref{fig:IRAS60100+CO.total.lb}. Figure \ref{fig:IRAS100+CO.loop+loc.lb} shows four panels of CO in different velocity intervals and suggests that both loops 1 and 3 contribute significantly to the main dust emission at $b$ above 1$^\circ$ and $l$ = 355$^\circ$ -- 358$^\circ$, in addition to the western half of loop 2 at $l$ $\sim$ 355.3$^\circ$. 

 We estimated the total hydrogen column density by combing the \atom{H}{}{}\emissiontype{I} and CO emissions (Figure \ref{fig:lb.NHdust.ps}a) and from the dust emission (Figure \ref{fig:lb.NHdust.ps}b). Here we describe the total column densities estimated by gas and dust as $N(\atom{H}{}{}\emissiontype{I})_{\mathrm{gas}}$ and $N(\atom{H}{}{}\emissiontype{I})_{\mathrm{dust}}$, respectively. We used the following relationship to convert the dust emission into $N$(\atom{H}{}{}\emissiontype{I})$_{\mathrm{dust}}$ by assuming the gas to dust mass ratio, $R_{\mathrm{gd}}$, of 100;
 \begin{eqnarray}
  N(\atom{H}{}{}\emissiontype{I})_{\mathrm{dust}} = \frac{4}{3}\left(\frac{a\rho}{Q_{\mathrm{100}}}\right)\frac{\tau_{100} R_{\mathrm{gd}}}{m_{\mathrm{H}}} \ (\mathrm{cm}^{-2})
 \end{eqnarray}
where $a$ is the grain radius, $\rho$ is the grain density in g cm$^{-3}$, $Q_{\mathrm{100}}$ is the grain emissivity at 100 $\mu$m, and $m_{\mathrm{H}}$ is the atomic hydrogen mass. We use an average value for $(a\rho/Q_{100})$ of 3.2 $\times$ (1000/$\lambda_{\mathrm{\mu m}}$)$^{-\beta}$, assuming $\beta = 2$, at wavelength $\lambda$ for a mixture of graphite and silicate grains (\cite{hil1983}; \cite{agl1996}). The optical depth at wavelength $\lambda$, $\tau_\lambda$, is represented as following;
 \begin{eqnarray}
 \tau_\lambda = \frac{f_\lambda}{B_\nu(\lambda,  T_{\mathrm{dust}})}
 \end{eqnarray}
where $f_\lambda$ is flux density at $\lambda$, and $B_\nu(\lambda, T_{\mathrm{dust}})$ is Planck function. We estimated dust temperature, $T_{\mathrm{dust}}$, by comparing flux densities of 60 $\mu$m and 100 $\mu$m at each observed point shown in Figure \ref{fig:lb.NHdust.ps} with following equation;
 \begin{eqnarray}
  \frac{f_{60}}{f_{100}} = \left(\frac{60}{100}\right)^{-3} \left(\frac{Q_{60}}{Q_{100}}\right)^\beta 
  \left[\frac{\exp(hc/\lambda_{100}k_bT_{\mathrm{dust}}) - 1}{\exp(hc/\lambda_{60}k_bT_{\mathrm{dust}}) - 1}\right] \frac{\Omega_{60}}{\Omega_{100}}
 \end{eqnarray}
where $h$ is Planck constant, and $c$ is the light speed, and $k_b$ is Boltzmann constant, and $\Omega_{60}$ and $\Omega_{100}$ are the solid angles at 60 $\mu$m and 100$\mu$m, respectively. We assume that $\Omega_{60}$ = $\Omega_{100}$. 
 
The relationship between $N$(\atom{H}{}{}\emissiontype{I})$_{\mathrm{gas}}$ and both the CO and \atom{H}{}{}\emissiontype{I} emissions is given as follows;
\begin{eqnarray}
N(\atom{H}{}{}\emissiontype{I})_{\mathrm{gas}} = (W(\atom{C}{}{}\atom{O}{}{}) \times 1.6 \times 10^{20} \times 2) + (W(\atom{H}{}{}\emissiontype{I}) \times 1.823 \times 10^{18}) \ (\mathrm{cm}^{-2})
\end{eqnarray}
where $W$ is the integrated intensity in K km s$^{-1}$ units. We use the conversion factor from $W$(CO) to the molecular hydrogen column density $N$(H$_2$), what we call "X-factor", of $1.6 \times 10^{20}$ cm$^{-2}$/(K km s$^{-1}$). We shall derive this value in the next subsection.

Figure \ref{fig:NH+H2.vs.N_dustH+H2} shows the scatter plot of the two values of $N$(\atom{H}{}{}\emissiontype{I}) for the region shown in Figure \ref{fig:lb.NHdust.ps}. The $N$(\atom{H}{}{}\emissiontype{I}) values show a good correlation with a correlation coefficient of log$_{10}$($N$(H)$_{\mathrm{dust}}$) = 0.76 $\times$ log$_{10}$($N$(H)$_{\mathrm{gas}}$) + 5.47 and support the physical association between the dust emission and the gaseous loops. 

\subsection{Mass estimate}

We made mass estimates of loops 1 and 2 by using three methods; i) molecular mass by using the X-factor newly derived from the present \atom{C}{}{13}\atom{O}{}{} and \atom{C}{}{12}\atom{O}{}{} ratio, ii) \atom{H}{}{}\emissiontype{I} mass, and iii) dust mass.  

Total molecular mass is estimated using the X-factor, which is an empirical conversion factor from \atom{C}{}{12}\atom{O}{}{} integrated intensity to molecular hydrogen column density:
 \begin{eqnarray}
  X = N(\mathrm{H}_{2}) / W(\atom{C}{}{12}\atom{O}{}{}) \ (\mathrm{cm^{-2}/(K km s^{-1})})
 \end{eqnarray}
Here we use the integration ranges same those in Figure \ref{fig:Loop1.lblv+13CO.2}; i.e., -180 to -90 km s$^{-1}$ for loop 1 and -90 to -40 km s$^{-1}$ for loop 2. The mass of the molecular gas is calculated using following equation:
 \begin{eqnarray}
	M(\mathrm{\atom{C}{}{12}\atom{O}{}{}}) = u m_{\mathrm{H}} \sum^{}_{} \bigl[ D^{2} \Omega N(\mathrm{H}_{2}) \bigr]
 \end{eqnarray}
where, $u$ is the mean molecular weight which is assumed to be 2.8, $D$ is the distance of loops 1 and 2, 8.5 kpc, and $\Omega$ is the solid angle subtended by a unit grid spacing 4$'$ $\times$ 4$'$. Two values of the X-factor, $0.24  \times 10^{20}$ cm$^{-2}$/(K km s$^{-1}$) \citep{oka1998} and $0.6 \times 10^{20}$ cm$^{-2}$ K km s$^{-1}$ cm$^{-2}$/(K km s$^{-1}$) \citep{oni2004} were previously derived in the Galactic center by using the virial mass. The dynamical state is however not in dynamical equilibrium when the magnetic flotation is operating and may be considerably different from the virial equilibrium. We therefore derive newly an X-factor by using the LTE mass derived from the present \atom{C}{}{13}\atom{O}{}{} and \atom{C}{}{12}\atom{O}{}{} data (Figure \ref{fig:Loop2.lblv+13CO.2}). 

	Although \atom{C}{}{13}\atom{O}{}{} observations are confined to $b <$ 1$^\circ$, \atom{C}{}{13}\atom{O}{}{} column densities are calculated by assuming the local thermodynamic equilibrium (LTE) condition to estimate the lower limit of molecular mass. The optical depth of \atom{C}{}{13}\atom{O}{}{}, $\tau(\mathrm{\atom{C}{}{13}\atom{O}{}{}})$, was calculated using following equation:
\begin{eqnarray}
       \tau(\mathrm{\atom{C}{}{13}\atom{O}{}{}}) = - \ln \left[ 1 - \frac{T^{*}_{\mathrm{R}}(\mathrm{\atom{C}{}{13}\atom{O}{}{}})}{5.29 \times (J(T_{\mathrm{ex}})-0.164)} \right]
 \end{eqnarray}
where, $T^{*}_{\mathrm{R}}(\mathrm{\atom{C}{}{13}\atom{O}{}{}})$ and $T_{\mathrm{ex}}$ are the radiation temperature and the excitation temperature of \atom{C}{}{13}\atom{O}{}{}, respectively. $J(T)$ is defined as $J(T) = 1 / \left[ \exp(5.29/T)-1 \right]$. $N(\mathrm{\atom{C}{}{13}\atom{O}{}{}})$ was estimated from:
\begin{eqnarray}
       N(\mathrm{\atom{C}{}{13}\atom{O}{}{}}) = 2.42 \times 10^{14} \frac{\tau(\mathrm{\atom{C}{}{13}\atom{O}{}{}}) \Delta V T_{\mathrm{ex}}}{1-\exp(-5.29/T_{\mathrm{ex}})} \ (\mathrm{cm}^{-2})
 \end{eqnarray}
where, $\Delta V$ is the \atom{C}{}{13}\atom{O}{}{} linewidth. $N(\mathrm{\atom{C}{}{13}\atom{O}{}{}})$ is calculated for each observed position which matches the observed position in \atom{C}{}{12}\atom{O}{}{}. In this study, We assume $T_{\mathrm{ex}}$ of 40 K, which is typical kinetic temperature estimated by a LVG analysis with multiple \atom{C}{}{}\atom{O}{}{} excitation lines in the foot point found $l \sim 356.0^\circ$ reported by Torii et al (2009b). Then, we adopt the ratio [H$_2$]/[\atom{C}{}{13}\atom{O}{}{}] = 10$^{-6}$ \citep{lis1989} to convert $N$(\atom{C}{}{13}\atom{O}{}{}) into $N$(H$_2$). The results are shown in Figure \ref{fig:xfactor.plot} which show that X-factor is estimated as $1.6 (\pm0.7) \times 10^{20}$ [cm$^{-2}$/(K km s$^{-1}$)]. This value is close to those derived from gamma ray observations for the inner disk $1.6 \times 10^{20}$ \citep{hun1997}, $1.9 \times 10^{20}$ \citep{str1996} and is an order of magnitude larger than those derived from the virial theorem.
By using this X-factor, the total molecular masses of loop 1 and 2 are calculated as $2.1 (\pm 0.9) \times 10^6 \MO,$ and $3.4 (\pm1.5) \times 10^6 \MO$, respectively.

\atom{H}{}{}\emissiontype{I} column densities are calculated assuming that the \atom{H}{}{}\emissiontype{I} 21 cm line is optically thin and $T_\mathrm{s} \gg h\nu/k$. The following equation is used:
 \begin{eqnarray}
	N(\atom{H}{}{}\emissiontype{I}) = 1.82 \times 10^{18} \times W(\atom{H}{}{}\emissiontype{I}) = 1.82 \times 10^{18} \int T^{*}_{\mathrm{R}} dv \ (\mathrm{cm}^{-2})
 \end{eqnarray}
The masses of the atomic gas of loops 1 and 2 are then derived as 2.4 $\times$ 10$^5$ $\MO$, including the mass of 0.7 $\times 10^5 \MO$ inside of the loop, and 2.5 $\times$ 10$^5$ $\MO$, respectively.
	
We also find that the dust emission likely corresponds to the loops at 60 $\mu$m and 100 $\mu$m of the IRAS data as shown in Figure \ref{fig:IRAS60100+CO.total.lb}. It is not certain what fraction of the IRAS emission is from loops 1 and 2 because there is another loop, loop 3, in the same direction. We find above that the total dust mass toward loops 1, 2 and 3 is well proportional to the dust mass as expressed by a linear relationship (Figure \ref{fig:NH+H2.vs.N_dustH+H2}) between the far infrared emission and the CO and \atom{H}{}{}\emissiontype{I} emissions, and the dust emission is well explained as associated with the molecular and atomic gas in the loops for the nominal gas to dust mass ratio of 100. Here we assume that the dust emission is proportional to $W(\atom{C}{}{12}\atom{O}{}{})$. $W(\atom{C}{}{12}\atom{O}{}{})$ of loops 1 and 2 are $2.4 \times 10^4$ K km s$^{-1}$ and $3.9 \times 10^4$ K km s$^{-1}$, respectively, and $W(\atom{C}{}{12}\atom{O}{}{})$ of loop 3 is $1.4 \times 10^5$ K km s$^{-1}$. The relation ship in Figure \ref{fig:NH+H2.vs.N_dustH+H2} then indicates the total dust masses of loops 1 and 2 is $2.1(\pm0.9) \times 10^4 \MO$ and $3.4(\pm1.5) \times 10^4 \MO$, respectively.

The all estimated masses for loops 1 and 2 are summarized in Table \ref{tab:mass}. The total masses of loops 1 and 2 are reached to $2.3(\pm0.9) \times 10^6 \MO$ and $3.7(\pm1.5) \times 10^6 \MO$, respectively, and if we assume the uniform velocity dispersion of 30 km s$^{-1}$ in the loops, the total kinetic energy of each loop amounts to $\sim$2--3 $\times 10^{52}$ erg.

\section{Discussion}
\subsection{Magnetic flotation}

\citet{par1966} presented the pioneering work on the Parker instability in order to explain cloud formation of the foot point of magnetic loop in the galaxy. \citet{mat1988} and \citet{hor1988} made detailed studies of the Parker instability analytically with a linear analysis and numerically including nonlinearity. Their results indicate that the fundamental parameters are the Alfv$\mathrm{\acute{e}}$n speed $V_{\mathrm{A}} = B / \sqrt{4\pi\rho} $ and the pressure scale height $H$. The height and length of a loop is given as a few times $H$ and several times $H$, respectively. A typical timescale is given by the ratio $H/V_{\mathrm{A}}$. The molecular and atomic gas is generally frozen-in to the magnetic field lines even with weak ionization at ionization degrees of 10$^{-8}$ -- 10$^{-7}$ and the magneto-hydrodynamical (MHD) description holds well in the interstellar space. An outstanding observational property of a magnetic flotation loop is that the falling down gas becomes often supersonic and forms shock fronts at the both ends of the loop, where gas density and velocity dispersion become enhanced, and these compressed regions appear as foot points of enhanced density \citep{mat1988}. 

Two molecular loops, loops 1 and 2, were discovered in the Galactic center and an interpretation was presented that loops 1 and 2 are created by the Parker instability in the magnetized nuclear disk with a field strength of $\sim$150 $\mu$G (Paper I). These authors estimated the field strength by assuming energy equi-partition between the turbulent gas motion and the magnetic field. They also made two-dimensional numerical simulations of loops 1 and 2 and showed that the two loops are successfully reproduced by the magnetic flotation mechanism. The gas inside the loop flows down to the disk under the influence of the disk gravity, forming a uniform velocity gradient along the loop. This down flowing gas collides with the disk and often forms shock fronts with enhanced turbulent motions. The two loops in fact show foot points on their both ends as is consistent with the theoretical prediction. It is likely that the gas in the foot point is heated and compressed by the shock fronts. Therefore, the magnetic flotation has a potential to offer a coherent explanation both on the large velocity dispersions and high temperatures of the nuclear gas disk. It is however to be noted that these simulations were local in the disk, not including the effects of rotation.
	
Subsequently, \citet{mac2009} carried out three-dimensional global numerical simulations of the nuclear gas disk and showed that the magneto-rotational instability coupled with the Parker instability works to create more than a few 100 loops over a 1kpc-radius nuclear disk, where the axially symmetric Miayamoto-Nagai potential \citep{miy1975} was adopted as the stellar gravitational field. \citet{mac2009} have shown that one-armed non-axisymmetric density pattern of m = 1 mode is developed in the nuclear disk. In the half of the azimuthal area of the nuclear disk having lower density, magnetic pressure tends to become stronger compared with gas pressure and prominent magnetic loops tend to be formed preferentially on this side of the disk rather than in the other half having higher density. So, the predicted asymmetry of the global distribution of the loops in fact seems to be consistent with the observations that about three-fourths of the dense molecular gas, CMZ, is distributed in the positive Galactic longitudes and that three loops are distributed in the negative Galactic longitudes. 

\subsection{Alternative explanations}

Supershells created by multiple massive stellar explosions may be a possible explanation alternative to the magnetic floatation. In the Galaxy supershells associated with molecular clouds have been studied in about ten regions; they includes the Carina Flare (\cite{fuk1999}; \cite{daw2008}), Gum nebula \citep{yam1999} and other molecular shells \citep{mat2001}. The typical sizes of these shells are from 100 to 500 pc and the total velocity dispersions are 20 km s$^{-1}$ corresponding to the total kinetic energy of $\sim10^{51}$ ergs. As derived in the previous section, the total kinetic energy of the loops is 10 times larger than this energy involved in molecular supershells and, more importantly, the velocity spans of the loops are from 50 to 100 km s$^{-1}$ in total, considerably larger than that of supershells. Considering the high pressure in the Galactic center region, we infer that supershells in the Galactic center may have even smaller velocity dispersions. In addition, we see no indications of the stellar remnants in the central regions of the loops that could have driven the shells. We therefore conclude that the loops are not explained by supershells. 

We shall here discuss the connection between the bar like gravitational potential and the magnetic flotation. The bar like gravitational potential is considered to be a viable mechanims to explain the radial motions of the gas in the central few 100 pc of the Galaxy \citep{bin1991}. Numerical simulations were used to find that the flow of gas in the Galactic center is dominated by a bar that has corotation at $r$ = 2.4 $\pm$ 0.5 kpc, which is viewed at an angle of 16 $\pm 2^\circ$ from its major axis \citep{bin1991}. From the structure of the \atom{H}{}{}\emissiontype{I} terminal velocity envelope, it is deduced that the central mass density scales as $\rho \propto r^{-1.75}$ out to at least about 1.2 kpc along the bar's major axis. Consequently, the circular velocity curve is rising significantly through the radius range where a naive analysis of the tangent velocity leads to a falling rotation curve. Obviously, the bar like potential is relevant to create large scale non-radial motions but is not necessarily exclusive in the magnetic flotation picture. In order to incorporate the effects of the bar like potential, it is required to develop a model which adopts the bar potential instead of the Miyamoto-Nagai potential in the global simulation of the magnetized gas disk. It is naturally expected that magnetic flotation loops are also created in such calculations because the basic physics remains the same in connection with the Parker instability. 

\subsection{Model fitting}

The numerical simulations by \citet{mac2009} show that the magnetic loops tend to be distributed an approximately constant radius from the center. \citet{fuj2009} derived the geometrical and kinematic properties of the newly discovered loop 3 by assuming the constant radius. We shall here estimate the geometrical and kinematical parameters such as radius, rotation and radial velocities. 

\begin{figure}
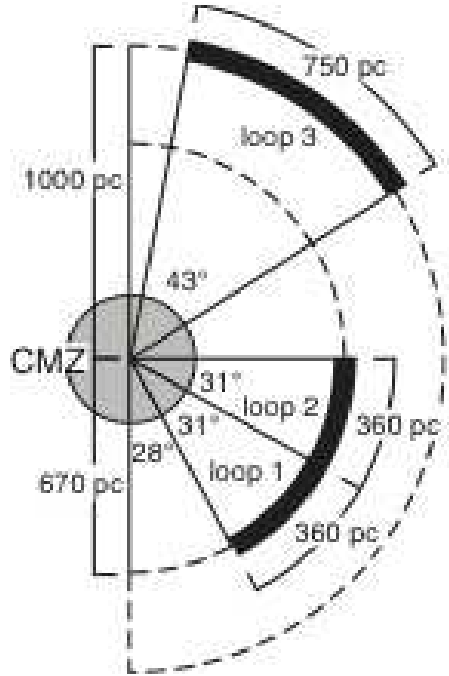

  \begin{center}
    \FigureFile(58.18mm,72.61mm){loop12.sch.eps2}
  \end{center}
  \caption{Schematic image of the position informations of loops 1, 2 and 3 with face on view. The parameters in the figure are also summarized in Table \ref{tab:pos}.}\label{fig:loop12.sch}
\end{figure}

First, we estimate the geometrical parameters of the loops. For simplicity we assume that the two loops have the same radius and physical length and shape and subtend the same angle with respect to the center. By considering triangles the corners of which are the Galactic center, the Sun, and a foot point of loops (the left foot point of loop 1: $l_0 \sim 357.5^\circ$, the right foot point of loop 1 and the left foot point of loop 2: $l_1 \sim 356.0^\circ$, and light foot point of loop 2: $l_2 \sim 355.5^\circ$), we can derive simultaneous equations by applying sine theorem to the three triangles;
\begin{eqnarray}
	&&R=\frac{R_{\mathrm{0}} \sin l_{\mathrm{0}}}{\sin (\theta_{\mathrm{0}} + l_{\mathrm{0}})} \\
	&&R=\frac{R_{\mathrm{0}} \sin (l_{\mathrm{1}})}{\sin (\theta_{\mathrm{0}} + \theta + l_{\mathrm{1}})} \\
	&&R=\frac{R_{\mathrm{0}} \sin (l_{\mathrm{2}})}{\sin (\theta_{\mathrm{0}} + 2\theta + l_{\mathrm{2}})}
\end{eqnarray}
where $R$ is the radius, and $R_0$ is the distance to the Galactic center, 8.5 kpc, and $\theta$ is the angles subtended by the arcs between the foot points at the Galactic center, and $\theta_0$ is offset. Schematic view of these parameters are shown in Figure \ref{fig:loop12.sch}.
We then solve the equations and derive $R$, $\theta_0$, and $\theta$ as $\sim$670 pc, $\sim$27.5$^{\circ}$, and $\sim$31$^{\circ}$, respectively. Therefore, the location of the left end of loop 1 is at $\theta_0$ $\sim$ 27.5$^\circ$, the right end of loop 1 and the left end of loop2 is at $\theta_0 + \theta \sim 58.5^\circ$, and the right end of loop 2 is at $\theta_0 + 2\theta \sim 89.5^\circ$. The projected length to the disk of each loops, $L$, is 360 pc, which is consistent with the results of \citet{tak2009}.

Next, we try to estimate the kinematical parameters of the loops. It is uncertain if the loops have radial motions. Some other features like the expanding molecular ring (EMR) are supposed to be expanding (\cite{saw2001}; \cite{mor1996}). We shall for simplicity assume that the loops are rotating and expanding from the center although the expansion is not yet confirmed in loops 1 and 2. We test the same method as discussed by \citet{fuj2009}. All the molecular gas of loop 1 and 2 has negative velocities. The velocity with respect to the LSR, $V_{\mathrm{LSR}}$, is then expressed as follows;
 \begin{eqnarray}
      V_{\mathrm{LSR}} &=&   V_{\mathrm{rot}} \frac{R_{0}}{R} \sin{l}
        -   V_{\mathrm{exp}}   \left( 1 - \frac{R_{0}^{2}}{R^{2}} \sin^{2}{l}  \right) ^{\frac{1}{2}}
        -   V_{\mathrm{sun}} \sin{l}
 \end{eqnarray}
where, $V_{\mathrm{rot}}$ is the rotational velocity of the disk, $V_{\mathrm{exp}}$ is the expansion velocity, and $V_{sun}$ is the rotation velocity of the LSR about the center of 200 km s$^{-1}$. We estimate that $V_{\mathrm{exp}}$ is 141 km s$^{-1}$ and $V_{\mathrm{rot}}$ is 47 km s$^{-1}$. These parameters are listed in Table \ref{tab:pos} for the three loops, and Figure \ref{fig:loop12.sch} shows schematic face-on view of these loops. The radius of loops 1 and 2 is somewhat smaller than that  of loop 3.

\subsection{Foot points; evidence for shock formation}

Foot points are characteristic features in the magnetic flotation loops. The general observational properties of the foot points are summarized as follows while the eastern foot point of loop1 seems to be under-developed compared with the other three;
1) The size and mass of the gas in the foot points are $\sim$20 pc and $\sim$10$^5$ $\MO$.
2) The spatial distribution shows a sharp intensity drop toward the plane
3) Velocity dispersions are as large as 50 -- 60 km s$^{-1}$. 
4) The line profiles are of triangle shape with some skewness. 

The two foot points in loop 2 seem quite consistent with the model prediction by \citet{mat1988} because they are located nearly at the same altitude of $\sim$ 150 pc above the galactic plane. The foot points also show strong intensity gradient toward the plane, suggesting shock fronts. The numerical simulations by \citet{mac2009} show that the large velocity dispersion is excited by the energy released by the falling-down gas. A LVG analysis of the \atom{C}{}{}\atom{O}{}{} $J$=1--0, 3--2, and 4--3 transitions indicates that the typical gas temperature is about 50 K in the foot points of loops 1 (Torii et al. 2009b). We also note that the high density in the foot points, $\sim10^3$ cm$^{-3}$ may offer a site of future star formation.

\subsection{Formation of molecular loops}

Generally speaking, it is easier for lower density gas to be lifted up rather than dense molecular gas by the magnetic flotation and it may seem odd that the dense molecular gas forms the loops. Observations suggest that such a loop may be initially formed within the disk as an elongated molecular gas tube and then rises up as observed, since the eastern part of loop 1 below the foot point may indicate such a case. 

Paper I made numerical simulations of loops for two-temperature, hot and cool, layers and showed that magnetic flotation works to form loops as observed. In order to reproduce the observed cool molecular loops in the Galactic center; \citet{tak2009} have carried out advanced two-dimensional numerical simulations for part of the disk containing the equatorial plane by adopting a stratified disk having three layers; cool layer ($T \sim 10^3$ K), warm layer ($T \sim 10^4$ K), and hot layer ($T \sim 10^5$ K). Their results show that the magnetic loops have a wavelength of $\sim$400 pc and a height of $\sim$350 pc. The typical time scale of the flotation given by the ratio $H/V_{\mathrm{A}}$ is 10 Myrs for $V_{\mathrm{A}}$ of 24 km s$^{-1}$ corresponding to density of 1000 cm$^{-3}$ and $B$ of 150 $\mu$G (Paper I). These values are fairly consistent with the observed parameters of loops 1 and 2; the length of the loops, 600 -- 800 pc (several times $H$), and the heights, 200 -- 300 pc (a few times $H$). In addition, it is important to note that their calculations produced the top-heavy loops which show the higher density layer at the top of the loops. This is because the loop in the cool layer has initially a small curvature, making the loop top flat, and the flat-top loops continuously rise to efficiently accumulate the gas around the loop top. Loop 1 actually shows that the higher density \atom{C}{}{}\atom{O}{}{} gas covers the lower density \atom{H}{}{}\emissiontype{I} is inside the CO loop (see Figure \ref{fig:LBLV.CO+HI}), indicating that the observations are well reproduced by the local numerical simulations by \citet{tak2009}.

Alternatively, the ambient \atom{H}{}{}\emissiontype{I} gas may be being converted into H$_2$ during the flotation. The shock compression by the rising loop in the upper part may lead to form molecular gas along the loop which is initially atomic. It is likely that the loop is surrounded by \atom{H}{}{}\emissiontype{I} gas, and the motion of the flotation at $V_{\mathrm{A}}$ inevitably causes shocks in the front side. This can lead to form H$_2$ in the shock compressed layer along the loop. The time scale of H$_2$ formation is given as 10$^9/n(H)$ yrs, where $n(H)$ is number density of \atom{H}{}{}\emissiontype{I} gas \citep{spi1978}, and is estimated to be 10 Myrs for $n(H)$ of 100 cm$^{-3}$. This time scale is consistent with the present ratio $H/V_{\mathrm{A}}$. It is also possible that \atom{H}{}{}\emissiontype{I} gas is rising continuously from the disk to the loop top as in the solar phenomena. The inside of loop 1 is in fact filled with lower intensity \atom{H}{}{}\emissiontype{I} gas. Higher resolution \atom{H}{}{}\emissiontype{I} observations may shed light on this possibility. It is shown by the TRACE satellite at UV that such continuous rising loops are actually observed in the sun.\footnote{http://trace.lmsal.com/} These actions can also accumulate molecular mass in the loop. \citet{tak2009} shows indeed that the flat top of the loops continuously rise and accumulates the gas efficiently. The higher density of the loop top derived by numerical results support this scenario.

\subsection{Magnetic field}

There is not yet direct measurements by the Zeeman effects of the magnetic field in the molecular clouds in the Galactic center region. \atom{O}{}{}\atom{H}{}{} measurements give only upper limits of several mG \citep{kil1992}. Observationally, it is a difficult task because the splitting is small and the \atom{O}{}{}\atom{H}{}{} spectra is so broad 100 km s$^{-1}$. 

Indirect arguments on the arched filaments suggest that the field strengths may be as large as 1 mG \citep{yus1987} and the double helical filaments may also suggest the strong magnetic field \citep{mor2006}. On the other hand, the strong field may not be occupying the whole volume as suggested by the low-frequency measurements of the field \citep{lar2006}.
The strong field from 100 $\mu$G to 1 mG is consistent with the strong stellar gravity of the central 1 kpc. The strong gravity results in differential rotation of the gas with frozen-in magnetic field and strong shearing motion is induced. This shearing motion amplifies the field. Thus, the strong field is a natural result of the strong stellar gravity.

\subsection{Comparisons with the Solar phenomena}

The present study revealed helical distributions of the loops. On the solar surface we find similar helical distributions which may be explained by MHD instabilities. Figures \ref{fig:helix_a} and \ref{fig:helix_b} reveal that loops 1 and 2 have helical distributions which change within a small velocity interval of $\sim$ 4 km s$^{-1}$. We shall compare these distributions with solar loops and theoretical calculations of magneto-hydrodynamics and discuss formation mechanisms of the helical distributions.

First, we tried to estimate the amplitude and wavelengths of the helix by applying sinusoidal and elliptical functions as follows;
 \begin{eqnarray}
     &&\left(
     \frac{\left(l-l_{\mathrm{c}}\right)-A\sin(k\Theta+\Delta\Theta)\cos\Theta}{X}
     \right)^2 \nonumber \\
     +
     &&\left(
     \frac{\left(b-b_{\mathrm{c}}\right)-A\sin(k\Theta+\Delta\Theta)\sin\Theta}{Y}
     \right)^2
     =1
 \end{eqnarray}
where $l_c$ and $b_c$ are the center position of an ellipse, $a$ and $b$ are the half lengths of two axes of ellipse, $A$ is amplitude of wave and $k$ is the number of waves in the ellipse, $\Theta$ is defined as $\Theta = arctan\left(Y(l-l_0)/X(b-b_0)\right)$, and $\Delta\Theta$ is a offset of $\Theta$.
As a result we find the parameters in Table \ref{tab:helix} show nice fits to the observations as shown in Figures \ref{fig:helix_a} and \ref{fig:helix_b}. The wavelengths of helix in loops 1 and 2 are estimated approximately 0.28$^\circ$(41 pc) -- 0.42$^\circ$(62 pc) and 0.28$^\circ$(41 pc), and the amplitudes are estimated approximately 0.2$^\circ$(29 pc) -- 0.34$^\circ$(50 pc) and 0.26$^\circ$(38 pc), respectively.

Next we compare with the solar phenomena. It has been known that solar prominences show helical distributions \citep{mat1998}. Formation of prominences are discussed under the two ideas below; 1) Magnetic flux tubes having helical shapes originally rise up. 2) Magnetic reconnection in the corona forms a helical shape. 

Recent Hinode observations show evidence for 1 \citep{oka2008} and we discuss along the idea 1 and compare the observations with numerical simulations. \citet{mat1998} show that distorted magnetic field lines becoming unstable against the kink instability form a helical shape while they rise. In fact soft X-ray observations of solar corona shows such a feature; \citet{wu2005} discussed if the helical distributions satisfy the condition for kink instability that the number of turns is large enough to cause the instability. The number of turns in the helical distrbutions in the top of the loops $\Phi = L_{\mathrm{loop}}B_\theta / rB_z$, where $L_{\mathrm{loop}}$ is the length of the loop, $r$ is the radius of the cross-section, $B_\theta$ and $B_z$ are the axial and azimuthal magnetic fields, respectively, are estimated as follows;\\
loop 1: $\Phi \sim 3\pi$\\
loop 2: $\Phi \sim 5\pi$\\
The both satisfies the condition for kink instability $\Phi > 2.5\pi$ \citep{hoo1981}, suggesting that the helical loops are subject to the kink instability of $m$=1. One possibility of formation of helical distributions is the gas in a rotating disk accumulated into a small volume by the Parker instability stores the magnetic distortions \citep{shi1991} and rise up to form the helical shape. The present findings of the helical distributions in loops 1 and 2 offers a support for that the magnetic floatation is a viable idea to explain these features and more detailed numerical simulations are desirable along this line.

\section{Conclusions}

We have presented a detailed study of magnetically floated loops 1 and 2 in the Galactic center based on the NANTEN \atom{C}{}{12}\atom{O}{}{} and \atom{C}{}{13}\atom{O}{}{}($J$=1--0) dataset.  The main conclusions are summarized below;

1) Molecular loops 1 and 2 have lengths of $\sim$300 -- 400 pc and heights of $\sim$200 -- 300 pc projected on the sky with a width of $\sim$30 pc. They are apparently connected with each other and have foot points of enhanced intensities and linewidths toward their both ends. The size and velocity span of the foot points are 20 -- 30 pc and 50 -- 100 km s$^{-1}$, respectively. Theoretical works on magnetically flotation loops predict formation of such a loop with the foot points as a result of falling-down motion to the disk along a loop and are consistent with the observational properties of the loops.

2) We show that the tops of loops 1 and 2 show helical distributions which may be due to magnetic instability. We also note that the current upper limits on the magnetic field strengths obtained in \atom{O}{}{}\atom{H}{}{} Zeeman measurements are consistent with the field strength required in the present picture $\sim$150 $\mu$G, while direct measurements of the Zeeman splitting yet remains to be achieved.

3) By assuming that the loops are of the same size at a radius $R$ from the center, we estimate the geometrical and kinematical properties of the loops such as the projected length of a loop, $L$, the angle subtended by a loop, $\theta$, and the rotation and expansion velocities, $V_{\mathrm{rot}}$ and $V_{\mathrm{exp}}$. The results are as follow; $R$ = 680 pc, $L$ = 350 pc, $\theta$ = 30$^\circ$ and $V_{\mathrm{rot}}$ = 47 km s$^{-1}$ and $V_{\mathrm{exp}}$ = 140 km s$^{-1}$. 

4) The two loops are associated with \atom{H}{}{}\emissiontype{I} gas, which basically shows similar distribution and kinematics with molecular gas. \atom{H}{}{}\emissiontype{I} gas of lower intensity is distributed inside the loops where \atom{C}{}{}\atom{O}{}{} emission is not detected. We also identified that the dust emission at 60 $\mu$m and 100 $\mu$m obtained with the IRAS is also associated with loop 1. The total mass of the loops are estimated by combining the molecular data, \atom{H}{}{}\emissiontype{I} data and the dust emission. The total mass of loops 1 and 2 including all these components are estimated to be 2.3($\pm$0.9) $\times$ 10$^6$ $\MO$ and 3.7($\pm$1.5) $\times$ 10$^6$ $\MO$. The mass of the molecular gas is much larger than the \atom{H}{}{}\emissiontype{I} mass, indicating that the atomic gas is a major component of the loops. The total kinetic energy of loops 1 and 2 are estimated to be $\sim$10$^{52}$ ergs for the observed velocity dispersions.

5) We have examined an alternative idea that supershells are responsible for the loop formation instead of the magnetic floatation. We find that the velocity spans of the loops, 50 -- 100 km s$^{-1}$, are significantly larger than the typical velocity dispersion of molecular supershells created by stellar explosions, 20 km s$^{-1}$, and that there are no known stellar objects inside the present loops. The velocity distribution which is dominated by a large monotonic velocity gradient is either not consistent with the kinematics expected for a spherical expansion. We thus conclude that supershells are not likely cause of the loops.

6) The recent global numerical simulations of magneto-hydrodynamics in the nuclear disk of 2-kpc radius by \citet{mac2009} indicate formation of the loops is a natural consequence of a differentially rotating magnetized gas disk under the strong stellar gravity. By adopting an axially symmetric gravitational potential, these authors find that the gas distribution tends to obey a $m$=1 mode where half of the disk has lower densities and that the well developed loops tend to be formed in this lower density half. In addition, two-dimensional local simulations are made by \citet{tak2009}, which assume stratified gas consisting of a cool layer ($T \sim 10^3$ K), warm layer ($T \sim 10^4$ K), and hot layer ($T \sim 10^5$ K) to confirm the observations. The results show that the flat top loops effectively accumulate the gas and make a dense top layer. We note that the properties of the loops shown by these simulations are consistent with the observations. We also discuss that the bar-like potential employed to explain the radial motions of the molecular gas is compatible with the magnetic flotation. Such a potential may be adopted as a next step in the global numerical simulations as a more realistic model.

 \bigskip
 
 We greatly appreciate the hospitality of all staff members
 of the Las Campanas Observatory of the Carnegie Institution
 of Washington. The NANTEN telescope was operated based
 on a mutual agreement between Nagoya University and the
 Carnegie Institution of Washington. We also acknowledge
 that the operation of NANTEN can be realized by contributions
 from many Japanese public donators and companies.
 This work is financially supported in part by a Grant-in-Aid
 for Scientific Research (KAKENHI) from the Ministry
 of Education, Culture, Sports, Science and Technology of
 Japan (Nos.~15071203 and 18026004) and from JSPS (Nos.~14102003, 20244014, and 18684003). 
 This work is also financially
 supported in part by core-to-core program of a Grant-in-Aid 
 for Scientific Research from the Ministry of Education,
 Culture, Sports, Science and Technology of Japan (No.~17004).
\newpage


\newpage

\begin{table}
\begin{center}
\caption{Observation summary}
  \begin{tabular}{cccccccc}
   \hline
   \hline
   \multirow{2}{*}{transition} & frequency & beamsize & grid & \multicolumn{2}{c}{observing region} & \multirow{2}{*}{total observed points} & mean r.m.s.\\
    & (GHz) & (') & (') & $l$ ($^{\circ}$) & $b$ ($^{\circ}$) &  & (K)\\
   \hline
   \atom{C}{}{12}\atom{O}{}{}($J$=1--0) & 115.27120 & 2.6 & 4 & -10 -- 10 & -5 -- 5 & 54000 & 0.3 \\
   \atom{C}{}{13}\atom{O}{}{}($J$=1--0) & 110.20137 & 2.6 & 2 & -6 -- 8 & -1 -- 1 & 25200 & 0.2 \\
      
   \hline
   \multicolumn{6}{@{}l@{}}{\hbox to 0pt{\parbox{110mm}{\footnotesize
   
   }\hss}} \label{tab:obs}
   \end{tabular}
\end{center}
\end{table}

\begin{table}
\begin{center}
\caption{Properties of the loops}
  \begin{tabular}{cccccccc}
   \hline
   \hline
   \multirow{2}{*}{Name} & $l$ & $b$ & V$_{\mathrm{LSR}}$  & M$_{\mathrm{CO}}$ & M$_{\atom{H}{}{}\emissiontype{I}}$ & M$_{\mathrm{dust}}$ \\
    & ($^\circ$) & ($^\circ$) & (km s$^{-1}$) & ($\times$ 10$^6\MO$) & ($\times$ 10$^6\MO$) & ($\times$ 10$^4\MO$) \\
   \hline
   
  loop 1 & 355.8 -- 358.0 & 0.5 - 1.6 & -180 -- -90 & 2.1$\pm$0.9 & 0.2 & 2.1$\pm$0.9 \\  
  loop 2 & 355.2 -- 356.6 & 0.6 - 2.2 & -90 -- -40   & 3.4$\pm$1.5 & 0.3 & 3.4$\pm$1.5 \\
  loop 3 & 355.0 -- 359.0 & 0.0 - 2.0 & 30 -- 160 & 12.0$\pm$5.2 & 2.1 & 14.1$\pm$5.2 \\  
      
   \hline
   \multicolumn{7}{@{}l@{}}{\hbox to 0pt{\parbox{120mm}{\footnotesize
     \par\noindent
   }\hss}} \label{tab:mass}
   \end{tabular}
\end{center}
\end{table}

\begin{table}
  \begin{center}
 \caption{Geometrical and kinematical parameters of loops 1 and 2}\label{tab:pos}
    \begin{tabular}{lcccccc}

\hline \hline 

\multirow{2}{*}{Name} & $R$ & $L$ & $\theta_0$ & $\theta$ & $V_{\mathrm{rot}}$ & $V_{\mathrm{exp}}$\\
				     & (pc) & (pc) & ($^\circ$) & ($^\circ$) & (km s$^{-1}$) & (km s$^{-1}$)\\
\hline

loops 1 \& 2 & 670 & 360 & 27.5 & 31.0 & 47 & 141\\
loop3           & 1000 & 750 &  129.5 & 43 & 80 & 130\\

\hline

    \end{tabular}
  \end{center}
\end{table}

\begin{table}
  \begin{center}
 \caption{Fitted parameters of helixes in loops 1 and 2}\label{tab:helix}
    \begin{tabular}{lccccccc}

\hline \hline 

\multirow{2}{*}{Name} & $l_0$ & $b_0$ & $X$ & $Y$ & $A$ & $k$ & $\Delta\Theta$\\
				     & ($^\circ$) & ($^\circ$) & ($^\circ$) & ($^\circ$) & ($^\circ$) &  & (rad.)\\

\hline

loop 1 \footnotemark[$\dagger$] & 353.40/353.40 & 0.55/0.47 & 4.0/4.0 & 1.7/1.7 & 0.17/0.10 & 30/24 & 0/0\\
loop 2 & 356.66 & -0.01 & 0.65 & 1.92 & 0.13 & 17 & 0.1\\

\hline
   \multicolumn{8}{@{}l@{}}{\hbox to 0pt{\parbox{120mm}{\footnotesize
     \par\noindent
  \footnotemark[$\dagger$] loop 1 shows two solutions. 
  }\hss}} 
    \end{tabular}
  \end{center}
\end{table}

\begin{figure}
  \begin{center}
    \FigureFile(110.7mm, 261.05mm){12CO.gcall.lb+lv.eps2}
  \end{center}
  \caption{Integrated intensity distributions of $^{12}$CO($J$=1--0) obtained by NANTEN. Dotted boxes in the figures show the regions that loops are located. (a) Velocity integrated intensity distributions. Integration range in velocity is from -300 to 300 km s$^{-1}$. Contours are plotted at 10, 20, 60, 100, 220, 340 K km s$^{-1}$ in black and 460, 820, 1180, 1540, 1900, 2260, 2620 K km s$^{-1}$ in white. (b) Longitude -- velocity diagram. Velocity resolutions are smoothed to 3 km s$^{-1}$. Integrated range in galactic latitude is from -2$^\circ$ to 2$^\circ$. Contours are plotted every 6 K degree from 0.25 K degree (black) and then every 1.2 K degree from 10 the one.}\label{fig:12CO.gcall.lb+lv}
\end{figure}

\begin{figure}
  \begin{center}
    \FigureFile(141.98mm,151.66mm){Loop1.lblv+13CO.2.eps2}
  \end{center}
  \caption{(a--d) Integrated intensity maps of loop 1 in \atom{C}{}{12}\atom{O}{}{}($J$=1--0) and \atom{C}{}{13}\atom{O}{}{}($J$=1--0). 
(a) Position-position diagram of loop 1 in \atom{C}{}{12}\atom{O}{}{}($J$=1--0). loop 1 is seen toward $l\sim 356^{\circ }- 358^{\circ }$ , $b\sim 0^{\circ }- 1.5^{\circ }$ . The integration range in velocity is from -180 to -90 km s$^{-1}$ . Contours are plotted every 10 K km s$^{-1}$ (black line) from the lowest one of 7.0 K km s$^{-1}$ ($3\sigma$) and every 20 K km s$^{-1}$ (white line) from the 6th one. (b) Position-position diagram in \atom{C}{}{13}\atom{O}{}{}($J$=1--0). The integration range is the same in figure (a). Contours are plotted every 4 K km s$^{-1}$ from 3.5 K km s$^{-1}$. (c) Velocity-galactic longitude diagram of loop 1. The velocity is smoothed at a resolution of 2.0 km s$^{-1}$. The integration range in the galactic latitude is shown by the solid lines in figure (a). Contours are ploted every 0.12 K deg from the lowest one of 0.20 K degree. An approximate position of the 3kpc arm is indicated by a dotted line. (d) Velocity-galactic longitude diagram of loop 2. The velocity is smoothed at a resolution of 2.0 km s$^{-1}$. The integration range is the same as figure (c). Contours are plotted every 0.05 K degree from 0.035 K degree.}\label{fig:Loop1.lblv+13CO.2}
\end{figure}

\begin{figure}
  \begin{center}
    \FigureFile(141.98mm,151.66mm){Loop2.lblv+13CO.2.eps2}
  \end{center}
  \contcaption{(e--h) Integrated intensity maps of loop 2 in \atom{C}{}{12}\atom{O}{}{}($J$=1--0) and \atom{C}{}{13}\atom{O}{}{}($J$=1--0). 
(e) Position-position diagram of loop 1 in \atom{C}{}{12}\atom{O}{}{}($J$=1--0). loop 2 is seen toward $l\sim 355^{\circ }- 356^{\circ }$ , $b\sim 1^{\circ }- 2^{\circ }$. The integration range in velocity is from -90 to -40 km s$^{-1}$ . Contours are plotted every 10 K km s$^{-1}$ (black line) from the lowest one of 7.0 K km s$^{-1}$ ($3\sigma$) and every 20 K km s$^{-1}$ (white line) from the 6th one. (f) Position-position diagram in \atom{C}{}{13}\atom{O}{}{}($J$=1--0). The integration range is the same in figure (e). Contours are plotted every 4 K km s$^{-1}$ from 4 K km s$^{-1}$. (g) Velocity-galactic longitude diagram of loop 2. The velocity is smoothed at a resolution of 2.0 km s$^{-1}$. The integration range in the galactic latitude is from 0.7$^{\circ}$ to 2.5$^{\circ}$. Contours are ploted every 0.12 K deg from the lowest one of 0.20 K degree. (h) Velocity-galactic longitude diagram of loop 2. The velocity is smoothed at a resolution of 2.0 km s$^{-1}$. The integration range is the same as figure (g). Contours are plotted every 0.05 K degree from 0.035 K degree.}
\label{fig:Loop2.lblv+13CO.2}
\end{figure}

\begin{figure}
  \begin{center}
    \FigureFile(118.71mm,118.87mm){specloop12.new.eps2}
  \end{center}
  \caption{Spectra of loops 1 and 2 in \atom{C}{}{12}\atom{O}{}{}($J$=1--0) with black line and \atom{C}{}{13}\atom{O}{}{}($J$=1--0) with blue line.}\label{fig:specloop12.new}
\end{figure}

\begin{figure}
  \begin{center}
    \FigureFile(120.00mm,228.21mm){loopallvel2.eps2}
  \end{center}
  \caption{The distributions of the peak \atom{C}{}{12}\atom{O}{}{}(J$=1--0$) velocity in loops 1 and 2.  
The peak velocity is determined by gaussian fitting to the averaged  
five spectra taken as a cross for each position. loops 1 and 2 are ploted 5.2 K km s$^{-1}$(Bold contours) and 45.2 K km s$^{-1}$(dashed line). Thin contours are ploted every 10 km s$^{-1}$ from -200 to -40 km s$^{-1}$(a), from -120 to 0 km s$^{-1}$(b).}\label{fig:loopallvel2}
\end{figure}

\begin{figure}
  \begin{center}
    \FigureFile(153.72mm,181.15mm){12co.channel.01.eps2}
  \end{center}
  \caption{Velocity channel maps of loops 1 and 2 in \atom{C}{}{12}\atom{O}{}{}($J$=1--0) integrated over successive 10 km/s. Contours are plotted every 4.8 K km/s (black contours) from 3.6 K km/s and every 9.6 K km/s (white contours) from 32.4 K km/s}\label{fig:12co.channel.01}
\end{figure}

\begin{figure}
  \begin{center}
    \FigureFile(153.72mm,181.15mm){12co.channel.02.eps2}
  \end{center}
  \contcaption{continued.}\label{fig:12co.channel.02}
\end{figure}

\begin{figure}
  \begin{center}
    \FigureFile(139.37mm,187.82mm){helix_b.eps2}
  \end{center}
  \caption{Velocity-integrated intensity map helical feature in loop 2. (a) Velocity integrated intensity map in \atom{C}{}{12}\atom{O}{}{}($J$=1--0). The integration range is from -86 to -66 km/s. Contours are plotted every 2 K km/s from 2.6 K km/s. Black box indicates the region shown in figure (b). (b) Velocity channel maps of helical feature integrated over 4 km/s with interval of 2 km/s. Contours are plotted every 0.8 K km/s from 1.6 K km/s. Dotted lines show the fitting result with the parameters shown in Table \ref{tab:helix}.}\label{fig:helix_b}
\end{figure}

\begin{figure}
  \begin{center}
    \FigureFile(112.65mm,183.45mm){helix_c.eps2}
  \end{center}
  \caption{Velocity-integrated intensity map helical feature in loop 2. (a) Velocity integrated intensity map in \atom{C}{}{12}\atom{O}{}{}($J$=1--0). The integration range is from -44 to -34 km/s. Contours are plotted every 2.4 K km/s from 2.6 K km/s. Black box indicates the region shown in figure (b). (b) Velocity channel maps of helical feature integrated over 4 km/s with interval of 2 km/s. Contours are plotted every 1.2 K km/s from 1.6 K km/s.}\label{fig:helix_c}
\end{figure}

\begin{figure}
  \begin{center}
    \FigureFile(177.49mm,148.8mm){LBLV.CO+HI.eps2}
  \end{center}
  \caption{Integrated intensity maps of \atom{H}{}{}\emissiontype{I} and \atom{C}{}{12}\atom{O}{}{}($J$=1--0) emission. The color image and dotted contours indicate \atom{H}{}{}\emissiontype{I} emission. Solid contours indicate \atom{C}{}{12}\atom{O}{}{}($J$=1--0) emission. (a) Velocity-integrated intensity maps of loop 1. Integration velocity range is from -180 to -90 km/s. \atom{H}{}{}\emissiontype{I}: Contours are plotted at 42 K km/s. \atom{C}{}{12}\atom{O}{}{}($J$=1--0): Contours are plotted every 4.8 K km/s from 3.6 K km/s and every 9.6 K km/s from 6th one. (b) Velocity-integrated intensity map of loop 2.  Integration velocity range is from -90 to -40 km/s. Contours are plotted at the same levels in figure (a). (c) Galactic longitude-velocity diagram of loop 1. The integrated galactic latitude range is shown in figure (a) by solid lines. \atom{H}{}{}\emissiontype{I}: Contours are plotted at 0.42 K degree. \atom{C}{}{12}\atom{O}{}{}($J$=1--0): Contours are plotted every 0.2 K degree from 0.12 K degree. (d) Galactic longitude-velocity diagram of loop 2. Contours are plotted at the same levels of figure (c).}\label{fig:LBLV.CO+HI}
\end{figure}

\begin{figure}
  \begin{center}
    \FigureFile(149.99mm,73.43mm){BV.CO+HI.eps2}
  \end{center}
  \caption{(a) Velocity-galactic latitude diagram of loops 1 and 2 in $^{12}$\atom{C}{}{}\atom{O}{}{}($J$=1--0) integrated from 355$^{\circ}$ to 358$^{\circ}$ in galactic longitude. Contours are plotted every 0.2 K degree from 0.24 K (black) degree and every 0.4 K degree from 9th one (white). (b) Velocity-galactic latitude diagram of loops 1 and 2 of \atom{H}{}{}\emissiontype{I} and \atom{C}{}{}\atom{O}{}{} integrated from 355$^{\circ}$ to 358$^{\circ}$ in galactic longitude. The color image and dotted contours indicate \atom{H}{}{}\emissiontype{I} emission. Solid contours indicate $^{12}$\atom{C}{}{}\atom{O}{}{}($J$=1--0) emission. \atom{H}{}{}\emissiontype{I}: Contous are plotted at 0.42 K degree. $^{12}$\atom{C}{}{}\atom{O}{}{}($J$=1--0): Contours are plotted at the same levels of figure (a).}\label{fig:BV.CO+HI}
\end{figure}

\begin{figure}
  \begin{center}
    \FigureFile(150.7mm, 161.05mm){IRAS60100+CO.total.lb.eps2}
  \end{center}
  \caption{CO Integrated intensity distributions superposed on IRAS 100$\mu$m. Integration range of CO is from -300 to 300 km s$^{-1}$. Contours are
plotted every 15 K km/s from 7 K km/s and every 30 K km/s from 5th one.}\label{fig:IRAS60100+CO.total.lb}
\end{figure}

\begin{figure}
  \begin{center}
    \FigureFile(150.7mm, 161.05mm){IRAS60100+HI.total.lb.eps2}
  \end{center}
  \caption{CO Integrated intensity distributions superposed on IRAS 60$\mu$m (left) and 100 $\mu$m (right). Integration range of CO is from -300 to 300 km s$^{-1}$. Contours are
plotted every  every 300 K km s$^{-1}$ from 100 K km s$^{-1}$.}\label{fig:IRAS60100+HI.total.lb}
\end{figure}

\begin{figure}
  \begin{center}
    \FigureFile(150.7mm, 161.05mm){IRAS100+CO.loop+loc.lb.eps2}
  \end{center}
  \caption{CO Integrated intensity distributions superposed on 100 $\mu$m. (a) Loop 1: Integration range is from -180 to -90 km s$^{-1}$. (b) Loop 2: Integration range is from -90 to -40 km s$^{-1}$. (c) Disk and local components: Integration range is from -40 to 30 km s$^{-1}$.  (d) Loop 3: Integration range is from 30 to 200 km s$^{-1}$. Contours in all figures are
plotted every 15 K km/s from 7 K km/s and every 30 K km/s from 5th one.}\label{fig:IRAS100+CO.loop+loc.lb}
\end{figure}

\begin{figure}
  \begin{center}
    \FigureFile(145.7mm, 161.05mm){lb.NHdust.ps.eps2}
  \end{center}
  \caption{(a) Spatial distributions of column density of atomic hydrogen, N(H), estimated from CO and \atom{H}{}{}\emissiontype{I} integrated intensity, N(H)$_{\mathrm{gas}}$, vs. N(H) estimated from IRAS 60 $\mu$m and 100 $\mu$m, N(H)$_{\mathrm{dust}}$, assuming dust-to-gas ratio of 100. Black boxes are the plotted regions in figure 8 in red.}\label{fig:lb.NHdust.ps}
\end{figure}

\begin{figure}
  \begin{center}
    \FigureFile(80.7mm, 101.05mm){NH+H2.vs.N_dustH+H2.eps2}
  \end{center}
  \caption{Scatter plot of the column density of atomic hydrogen, N(H), estimated from CO and \atom{H}{}{}\emissiontype{I} integrated intensity, N(H)$_{\mathrm{gas}}$, vs. N(H) estimated from IRAS 60 $\mu$m and 100 $\mu$m, N(H)$_{\mathrm{dust}}$, assuming dust-to-gas ratio of 100. Plotted region is the same shown in figure 7. Red points show the loop top regions shown in figure 7. Filled triangles denote the foot point of loops 1 and 2, and filled triangles and diamonds denote the top regions of loops 1 and 3 and top of loop 2, respectively. Dotted line show the fitting result for all points; log$_{10}$(N(H)$_{\mathrm{dust}}$) = 0.76 $\times$ log$_{10}$(N(H)$_{\mathrm{gas}}$) + 5.47}\label{fig:NH+H2.vs.N_dustH+H2}
\end{figure}

\begin{figure}
  \begin{center}
    \FigureFile(80.18mm,100.61mm){xfactor.plot.eps2}
  \end{center}
  \caption{Scatter plot of the integrated intensity, W(CO), vs. the column density of molecular hydrogen, N(H$_2$), estimated from \atom{C}{}{13}\atom{O}{}{}. Solid line show the fitting result; N(H$_2$) = 1.6 $\times$ 10$^{20}$ $\times$ W(\atom{C}{}{12}\atom{O}{}{}).}\label{fig:xfactor.plot}
\end{figure}

\end{document}